\documentstyle[graphicx,aps,prl,psfig]{revtex}
\begin{document}
\draft
\twocolumn[\hsize\textwidth\columnwidth\hsize\csname
@twocolumnfalse\endcsname
\title{Using of Phenomenological Piecewise Continuous Map\\
for Modeling of Neurons Behaviour}
\author{K.V. Andreev and L.V. Krasichkov}
\address{Department of Nonlinear Processes, Saratov State University,
83~Astrakhanskaya, 410026~Saratov, Russia}
\date{\today}
\maketitle
\begin{abstract}
A piecewise continuous map for modeling bursting and spiking 
behaviour of isolated neuron is proposed. The map was created 
from phenomenological viewpoint. The map demonstrates oscillations, 
which are qualitatively similar to oscillations generating by
Rose--Hindmarsh model. The synchronization in small ensembles of 
the maps is investigated. It is considered the different 
number of elements in the ensemble and different connectivity topologies.

\end{abstract}
\pacs{PACS number(s): 05.45.Xt, 87.17.Nn}

\narrowtext
\vskip1pc]

\section{Introduction}

Investigations of neuron ensembles (Central Pattern Generators, CPG) from 
the nonlinear dynamics viewpoint attract attention of many physicists last 
years. It is known some successful experiments with biological neurons (see, 
for example, \cite{r1}). 
Now exist a lot of mathematical models describing neuron 
behavior on the basis of ordinary differential equations \cite{r2}. 
Even simple 
models such as well-known FitzHugh--Nagumo and Rose--Hindmarsh systems, 
which may describe some important facts of neuron dynamics, need a lot of 
resources in numerical experiments for rather large ensembles. In this case 
the models with discrete time (maps) are much more appropriate.

It is known a set of such kind maps (i) the model with variable taking the 
discrete set of values (finite automata) \cite{r3,r4}, 
(ii) the map taken on the 
flow of differential equations system \cite{r5}, 
(iii) learning globally coupled 
excitable map system \cite{r6}, 
(iv) recently proposed two-dimensional map \cite{r7}.
However, the variety of models and methods for the neuron 
dynamics descriptions testifies that nonexists the common approach for
creation of maps defining the isolated neuron behavior as well as for
creation of the model neuron ensembles with pregiven topology and coupling
kind.    

In this paper is proposed a map with variable that changes continuously in 
specified range. This map demonstrates behavior that is qualitatively 
similar to the real neuron dynamics. It will be shown below that proposed 
map describes the complete synchronization in neuron ensembles. It will be 
also observed dependence the synchronization degree of the network 
configuration. Another results that will be discussed connected with the 
influence of external force on studied ensembles.

\section{The single neuron model}

\begin{figure}[htbp]
\centerline{\includegraphics[width=1.5in]{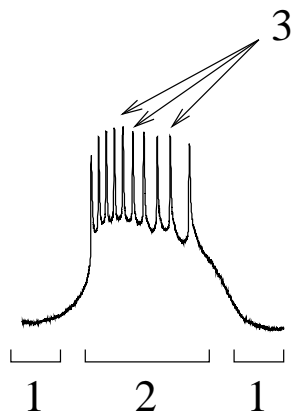}}
\vspace{0.2cm}
\caption{The fragment of time series of isolated biological 
neuron ("1" is the rest state region, 
"2" is the burst and "3" is the spikes).}
\label{fig1}
\end{figure}

The map was constructed on the assumption of phenomenological conceptions. 
The basic idea is founded on the fact that in neuron behavior at time series 
the three regions (rest, burst and spike regions) could be identified. This 
relative division is schematically shown in Fig.~\ref{fig1}.

The map ($x \rightarrow x'$) consists of two piecewise continuous 
functions connected by 
transition conditions. The system dynamics is described with four state 
variables: $x \in [0, 1]$, $d = \{-1, 1\}$, 
$s_{1}, s_{2} = \{ 0, 1\}$. The main 
``observable'' variable $x$  
qualitatively corresponds to membrane potential of 
the neuron, the variable $d$ is used to select 
one of two branches of function and the variables
$s_{1}, s_{2}$ are ``switches'', which define the conditions for burst
ending. 
In mathematical notation the map are written in the following way.

If $d=1$ (the increasing of the $x$ value)
\begin{equation}
\label{eq1}
{x}' = {\left\{ {\begin{array}{lll}
 \alpha_{1} \arctan(k_{1} x), &\mbox{if}& 
   x \in \left[ 0, a - \delta_{1} \right),  \\ 
 2a - x,  &\mbox{if}& x \in \left[ a - \delta_{1}, a \right), \\ 
 \gamma_{1} \left( x - a \right) + a, &\mbox{if}&
   x \in \left[ a, c \right), \\ 
 \end{array}} \right.}
\end{equation}

The transition condition: $d=-1$ if $x \in [c, 1]$.

If $d=-1$ (the decreasing of the $x$ value)
\begin{equation}
\label{eq2}
{x}' = {\left\{ {\begin{array}{lll}
 \frac{1}{\gamma_{2}} \left( x - a \right) + a, &\mbox{if}&
  x \in {\left[ a + \delta_{2}, 1 \right]}, \\ 
 2a - x, &\mbox{if}& 
  x \in \left[ a, a + \delta_{2} \right), s_{1} s_{2} = 1 \\ 
 \frac{1}{\alpha_{2}} \arctan(k_{2} x), &\mbox{if}&
  x \in \left[ \delta_{3}, a \right) . \\ 
 \end{array}} \right.}
\end{equation}

Additional terms: $d=1$ if $x \in \left[ a, a + \delta_{2} \right)$ and 
$s_{1} s_{2} = 0$,
\[
s_{1} = {\left\{ {\begin{array}{lll}
 1, &\mbox{if}& x \in {\left[ c, h_{1}  \right]}, \\ 
 0, &\mbox{if}& x \in {\left[ 0, a \right]}, \\ 
 \end{array}} \right.}
\quad
s_{2} = {\left\{ {\begin{array}{lll}
 1, &\mbox{if}& x \in {\left[ {h_{2}, 1} \right]}, \\ 
 0, &\mbox{if}& x \in {\left[ {0, a} \right]}, \\ 
 \end{array}} \right.}
\]
$d=1$ if $x \in \left[ 0, \delta_{3} \right)$.

Here $a$, $k_{1}$, $k_{2}$, $\gamma_{1}$, $\gamma_{2}$, $\delta_{1}$, 
$\delta_{2}$, $\delta_{3}$ are some constant parameters, and others 
coefficients are determined from continuity condition 
($c=(1-a)/\gamma_{1}+a, 
\alpha_{1}=a/\arctan(k_{1} a)$, 
$\alpha_{2}=a/\arctan(k_{2} a)$). Parameter 
$h_{2}$ is also constant and close to unity, and $h_{1}$ should depend on 
the value of $c$, that is $h_{1}=c+\Delta h$, where 
$\Delta h$ is the relatively small parameter.

\begin{figure}[htbp]
\centerline{\includegraphics[width=2.44in]{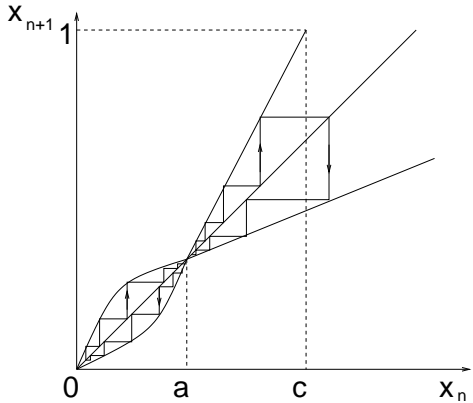}}
\vspace{0.2cm}
\caption{The iteration diagram for the map, describing by (\ref{eq1}),
(\ref{eq2}) transition conditions.}
\label{fig2}
\end{figure}

\begin{figure}[htbp]
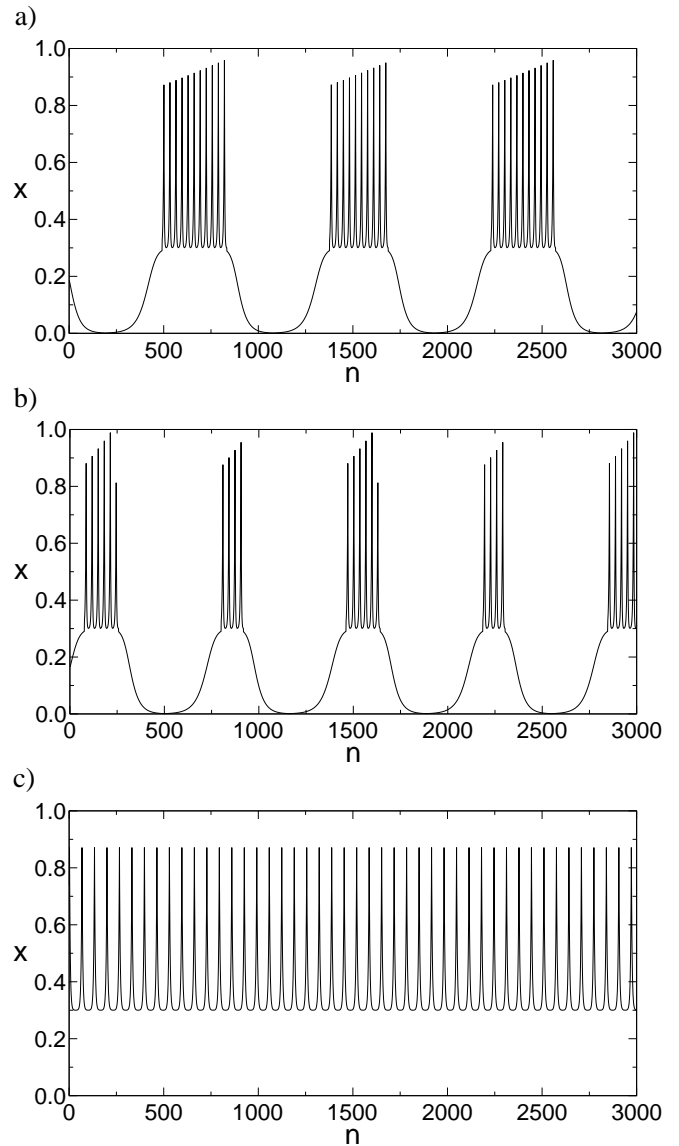

\centerline{\includegraphics[width=\columnwidth]{fig3a.eps}}
\centerline{\includegraphics[width=\columnwidth]{fig3b.eps}}
\centerline{\includegraphics[width=\columnwidth]{fig3c.eps}}
\vspace{0.2cm}
\caption{Characteristic time series for isolated model neuron for 
$\delta_2=0.001$, $h_1=0.88$, $\gamma_1=1.4$, $\gamma_2=1.75$ (a);
$\delta_2=0.001$, $h_1=0.88$, $\gamma_1=1.402$, $\gamma_2=1.75$ (b); 
$\delta_2=0.0001$, $h_1=0.92$, $\gamma_1=1.3$, $\gamma_2 = 1.3$ (c). The
values of the other parameters are $a=0.3$, $k_1=0.9$, $k_2=1.0$, 
$h_2=0.95$, $\delta_1=0.01$, $\delta_3 = 0.001$.}
\label{fig3}
\end{figure}

For better understanding how this map works the iteration diagram is shown 
in the Fig.~\ref{fig2}. Fig.~\ref{fig3} demonstrates the time series for 
the isolated model neuron 
described by the map for various values of control parameters.

\section{Generalization of the model to coupled neurons. Two coupled neurons}

Investigation of single neuron behavior has a little practical sense. Much 
more interesting might be model of neuron ensemble. Therefore this piecewise 
continuous map was generalized to coupled neuron systems. In case of the 
ensemble with $N$ elements with state variables 
$\left({x_{n}^{1}, x_{n}^{2}, ..., x_{n}^{N}} \right)$ 
the influence of neighbours to $j$-th neuron at 
($n+1$) discrete time is given by adding the following term to $x_{n+1}^{j}$ 
variable
\begin{equation}
\label{eq3}
{\frac{{1}}{{L_{j}}} }{\sum\limits_{\begin{array}{l}
{i = 1} \\ 
{i \ne j} \\ 
\end{array}}^{N} {\varepsilon _{ij} \left( {x_{n}^{i} - x_{n}^{j}}  \right) 
\Theta \left( {x_{n}^{i} - a} \right)}} ,
\end{equation}
where $\varepsilon_{ij}$, ($i,j=1, ..., N$) is the  connection weight between 
$i$-th and $j$-th elements, 
$L_{j}$ is the number of neighbours of $j$-th neuron, $\Theta$ is the  
Heaviside step function (authors presume that exists a threshold of 
interaction).

The first investigated small 
ensemble was constructed of two coupled neurons. The 
main interest of the system research was consisted in revealing complete 
synchronization. The degree of synchronization was determined as
\begin{equation}
\label{eq4}
\Delta = \frac{1}{M}{\sum\limits_{n = 1}^{M} 
{\left| x_{n}^{1} - x_{n}^{2}  \right|}},
\end{equation}
where $M$ is the number of iterations for a discrete time average.

\begin{figure}[htbp]
\centerline{\includegraphics[width=3.0in,angle=0]{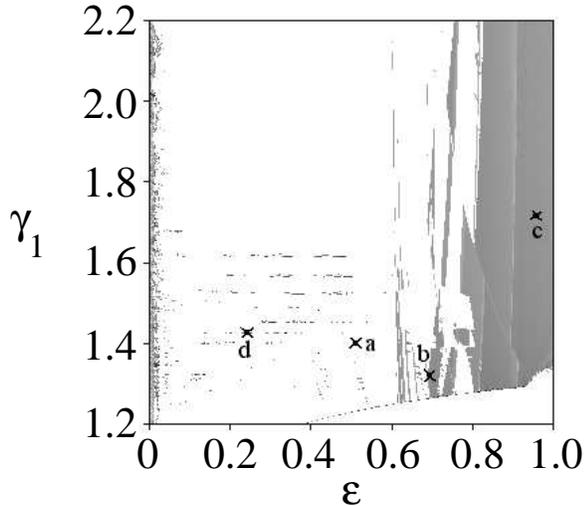}}
\vspace{0.2cm}
\caption{The parameter plane, plotted for two coupled model neurons. 
The letters
correspond to time series, presented in Fig.~\ref{fig5}. 
The values of the other
parameters are $a=0.3$, $k_{1}=0.9$, $k_{2}=1.0$, $\gamma_2=1.75$, 
$\delta_{1}=0.01$, $\delta_{2}=0.001$, $\delta_{3}=0.001$, 
$h_{2}=0.95$.}
\label{fig4}
\end{figure}

\begin{figure}[htbp]
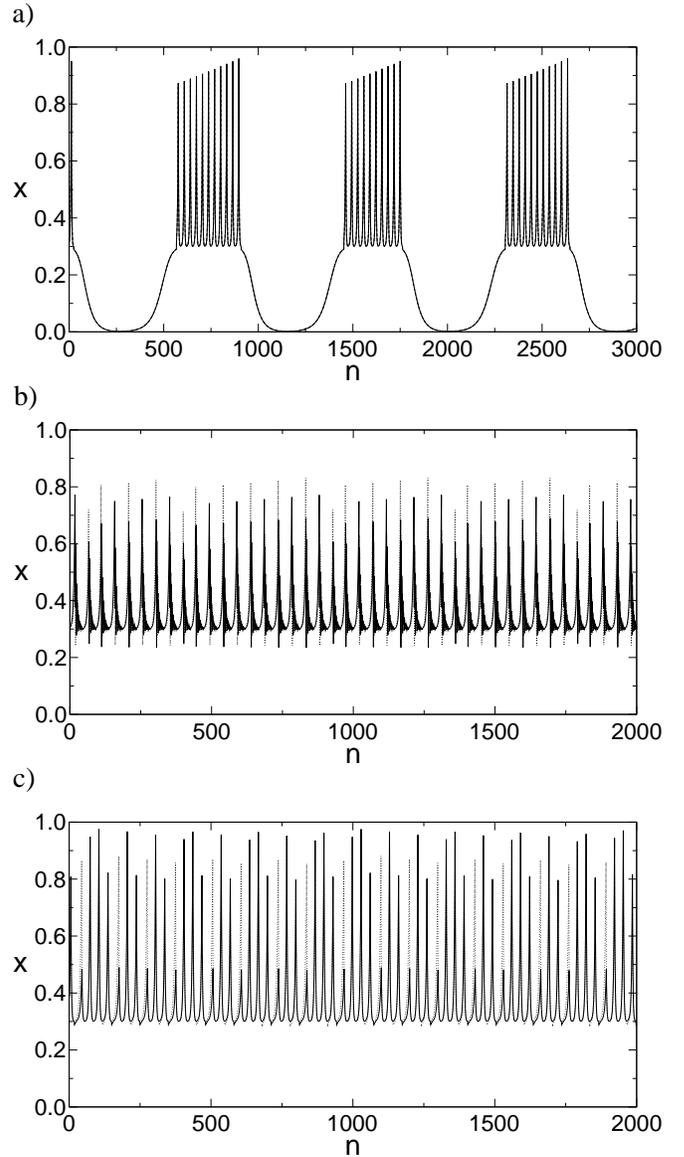

\centerline{\includegraphics[width=\columnwidth]{fig5a.eps}}
\centerline{\includegraphics[width=\columnwidth]{fig5b.eps}}
\centerline{\includegraphics[width=\columnwidth]{fig5c.eps}}
\vspace{0.2cm}
\caption{Characteristic time series plotted for two coupled neurons 
in the points, marked on the parameter plane in fig.~\ref{fig4} 
with $\varepsilon=0.50$, 
$\gamma_{1}=1.40$ (complete synchronization) (a); 
$\varepsilon=0.95$, $\gamma_{1}=1.70$ (b); 
$\varepsilon=0.24$, $\gamma_{1} = 1.427$ (c).}
\label{fig5}
\end{figure}

Value $\Delta = 0$ corresponds to the highest degree of synchronization, 
$\Delta > 0$ means that the synchronization is absent or 
synchronization is incomplete. In 
this research the degree of synchronization (\ref{eq4}) 
on the plane of parameters 
$\left( \varepsilon, \gamma_{1} \right)$ is  calculated. The choice of 
parameter $\varepsilon $ as a control parameter 
is natural because it's value 
determines the strength of connection. The value of parameter $\gamma_{1}$ 
(as well as $\gamma_{2}$) influences considerably to the system behavior, 
so it is chosen as second control parameter. The parameter planes are 
plotted in grayscale, namely, synchronization (for value $\Delta=0$) region  
marked with the white color, the positive value of $\Delta$ is 
subjected to the simple rule: the larger value, the darker point. Such 
parameter plane for two coupled neurons is presented in Fig.~\ref{fig4}. 
The behavior 
of this system in some points, marked on the parameter plane with letters, 
is shown in Fig.~\ref{fig5}.

It was revealed that the system dynamics depends on the initial conditions 
therefore the question about typicalness of presented parameter plane 
occurs. This problem was studied in detail. The view of the attraction 
basins, plotted in different points of the parameter plane, and the 
parameter planes obtained for various initial conditions shows that the 
general features of the planes are invariable, so this question will be not 
discussed below.

\section{Systems with several elements}

In the research is observed a large quantity of neuron ensembles with 
different numbers of elements and various spatial configurations. Some 
revealed regularities for ensemble of seven coupled neurons is 
presented here.

In case of several elements the degree of synchronization can't be defined 
with (\ref{eq4}), therefore in numerical experiments is used the following 
approximate relationship
\[
\Delta = \frac{1}{M} {\sum\limits_{n = 1}^{M} 
\left| {x_{n}^{k} - \frac{1}{N} 
{\sum\limits_{i = 1}^{N} {x_{n}^{i}}} }  \right|}.
\]

\begin{figure}[htbp]
\hbox{
\centerline{
\includegraphics[width=1.6in,angle=0]{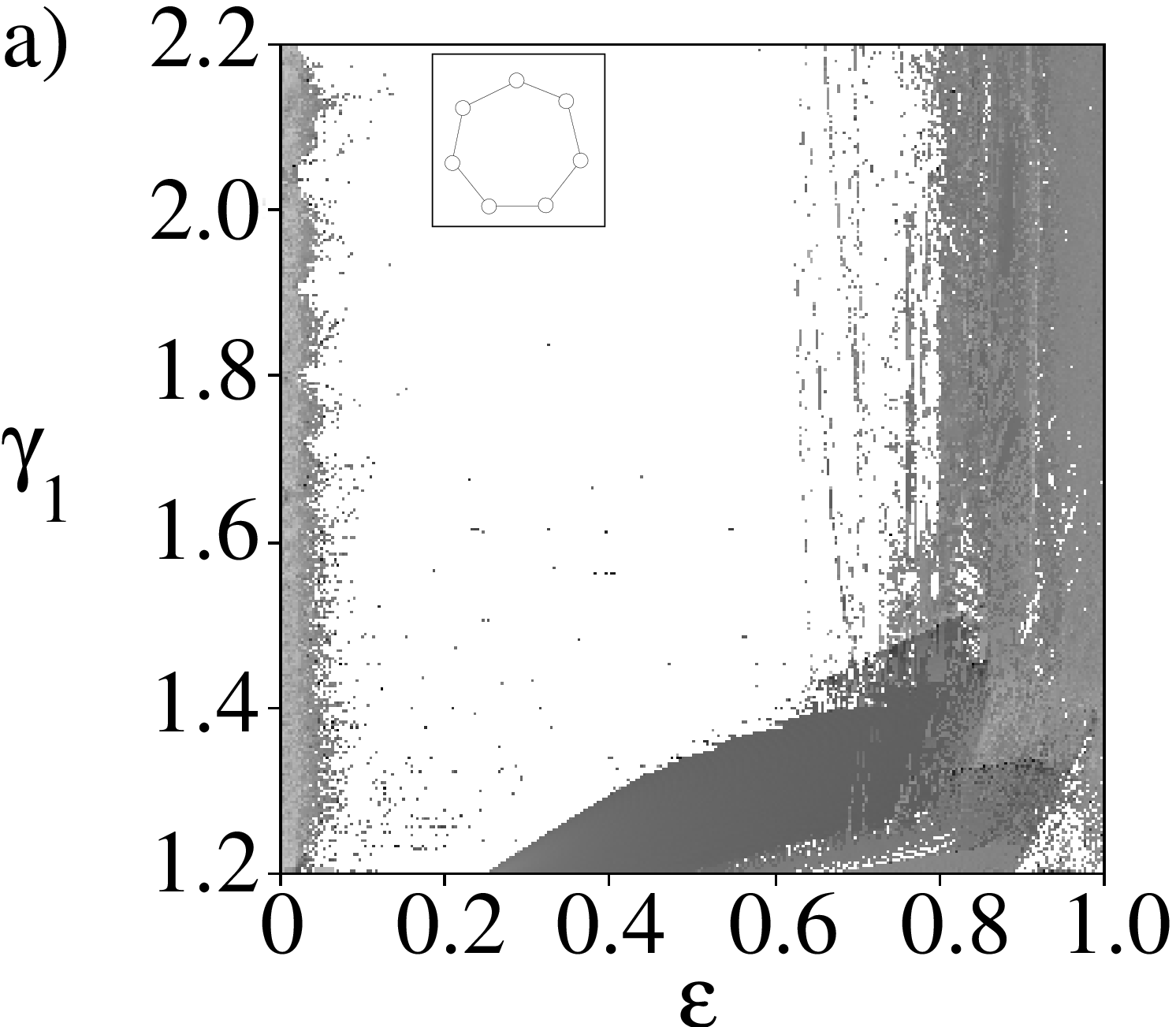}
\includegraphics[width=1.6in,angle=0]{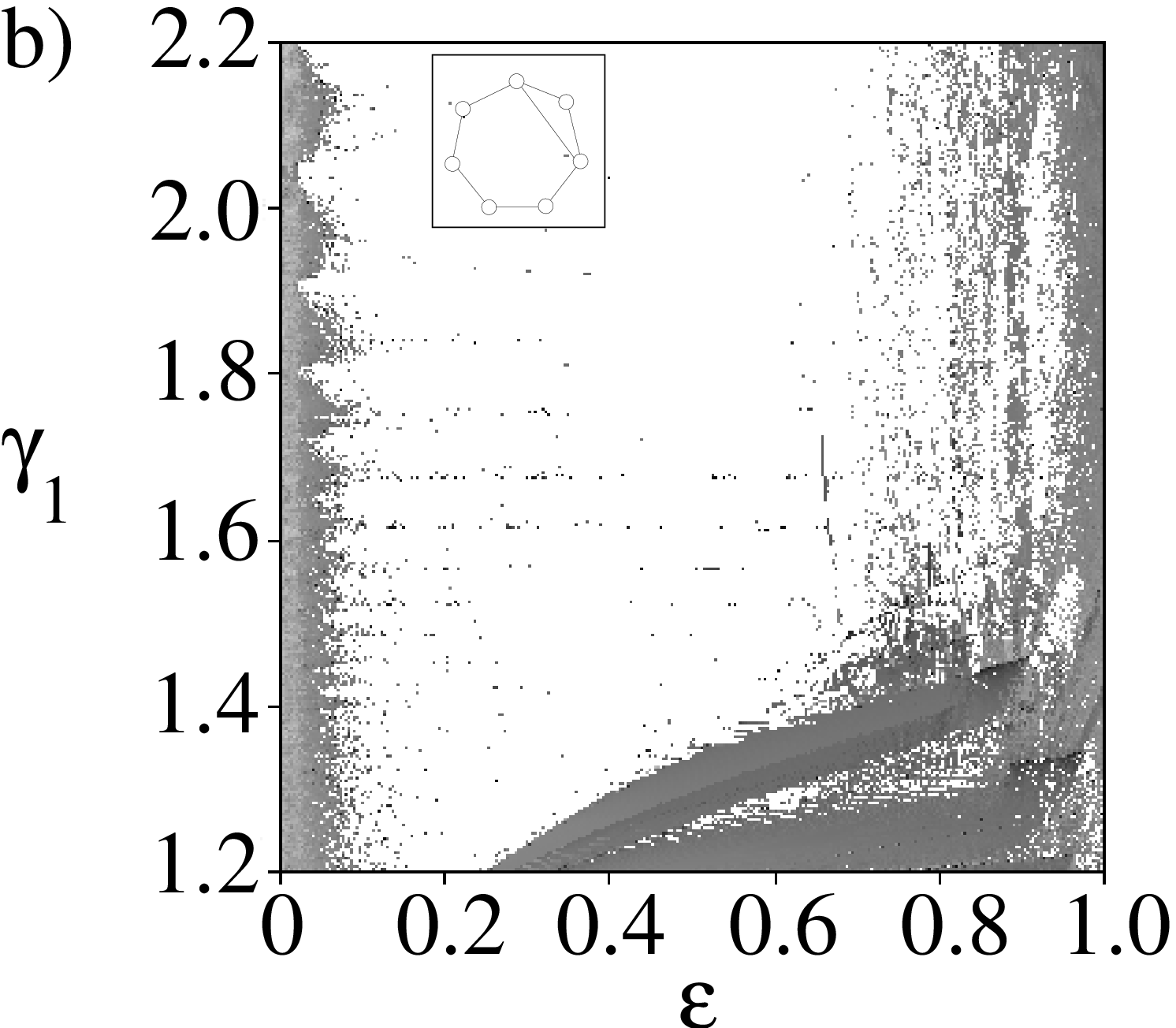}}}
\hbox{
\centerline{
\includegraphics[width=1.6in,angle=0]{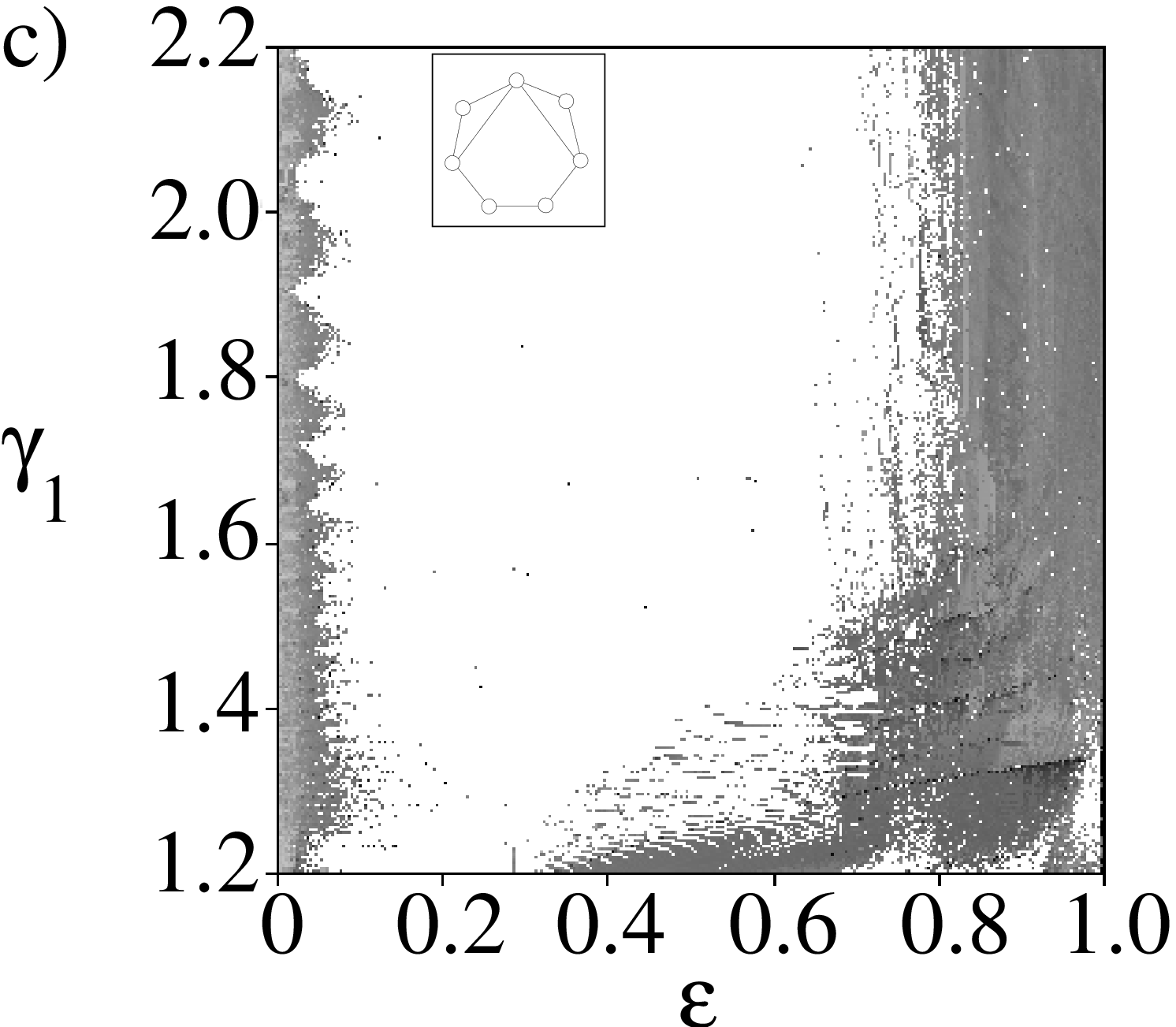}
\includegraphics[width=1.6in,angle=0]{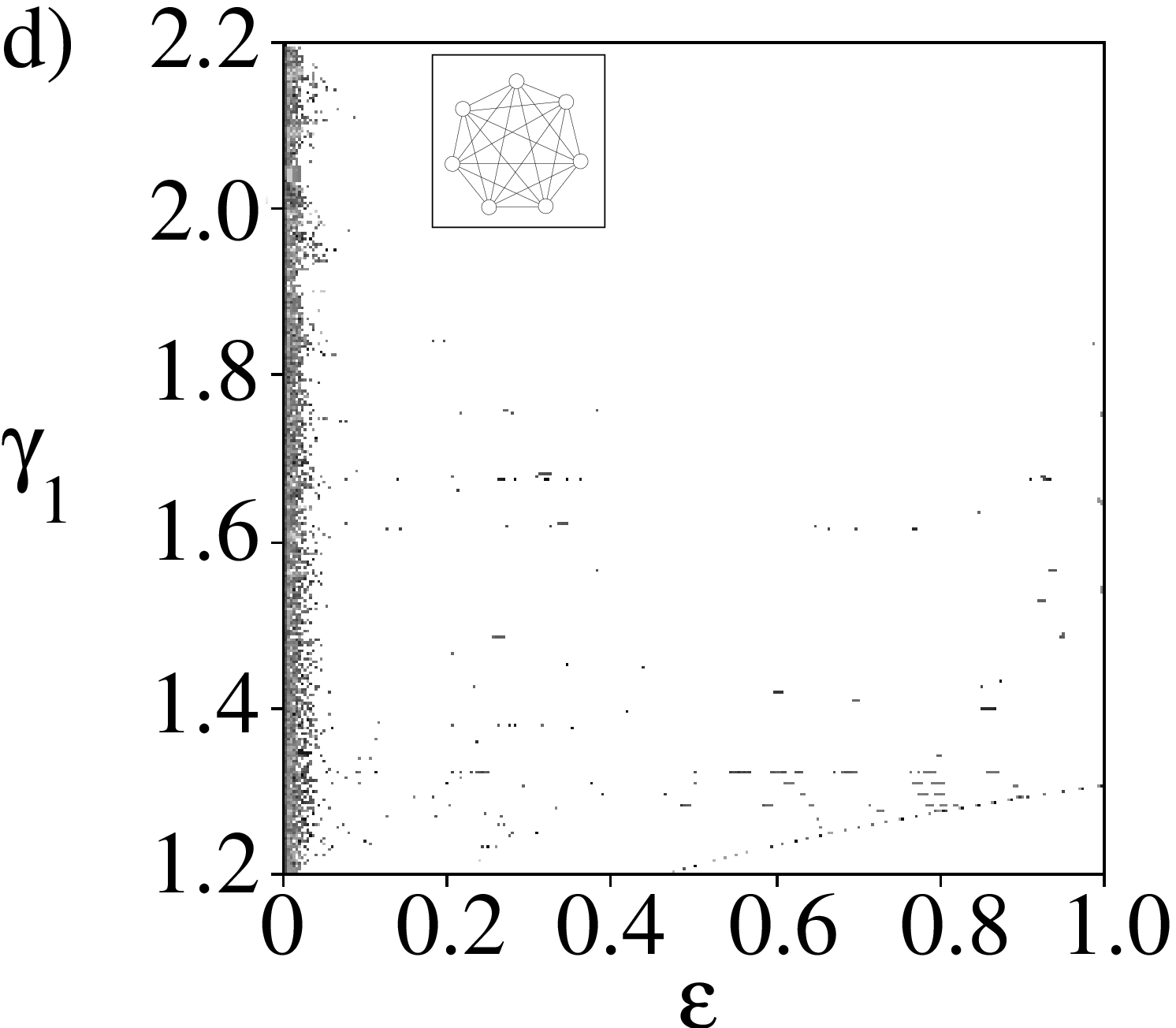}}}
\vspace{0.2cm}
\caption{The planes of parameter for ensembles with seven elements and
different spatial configurations (shown at the insets) 
obtained from ring-type system by adding of connections.}
\label{fig6}
\end{figure}

Fig.~\ref{fig6} presents the parameter planes that 
are gotten for systems 
schematically shown at the insets. As it can be seen, 
the areas with synchronous 
behaviour become larger when additional connections are inserted. 
Note, that although this tendency is clearly visible from Fig.~\ref{fig6},
the synchronous behaviour is observed in another areas of the parameter
plane (see Fig.~\ref{fig6}b and Fig.~\ref{fig6}c). Comperison of
Fig.~\ref{fig6}b and Fig.~\ref{fig6}c shows that synchronization degree is
decreased in the region 
$\varepsilon \in [0.8;~1.0]$ for all values $\gamma_1$. 
This result remains correct for systems with another numbers of elements.

It was  shown that the synchronization degree depends on the fast motion (the 
spikes region) and practically is independent of slow motions. This fact was 
revealed when the nonlinear branches (see Fig.~\ref{fig2}, region 
$x_n \in [0;~a]$) 
of the map functions (\ref{eq1}), (\ref{eq2}) were 
approximated with piecewise linear functions. There is the visible 
changes of the time series, but 
the view of parameter planes remains without any 
considerable changes. 

\section{The influence of the external action}

\begin{figure}[htbp]
\hbox{
\centerline{
\includegraphics[width=1.6in,angle=0]{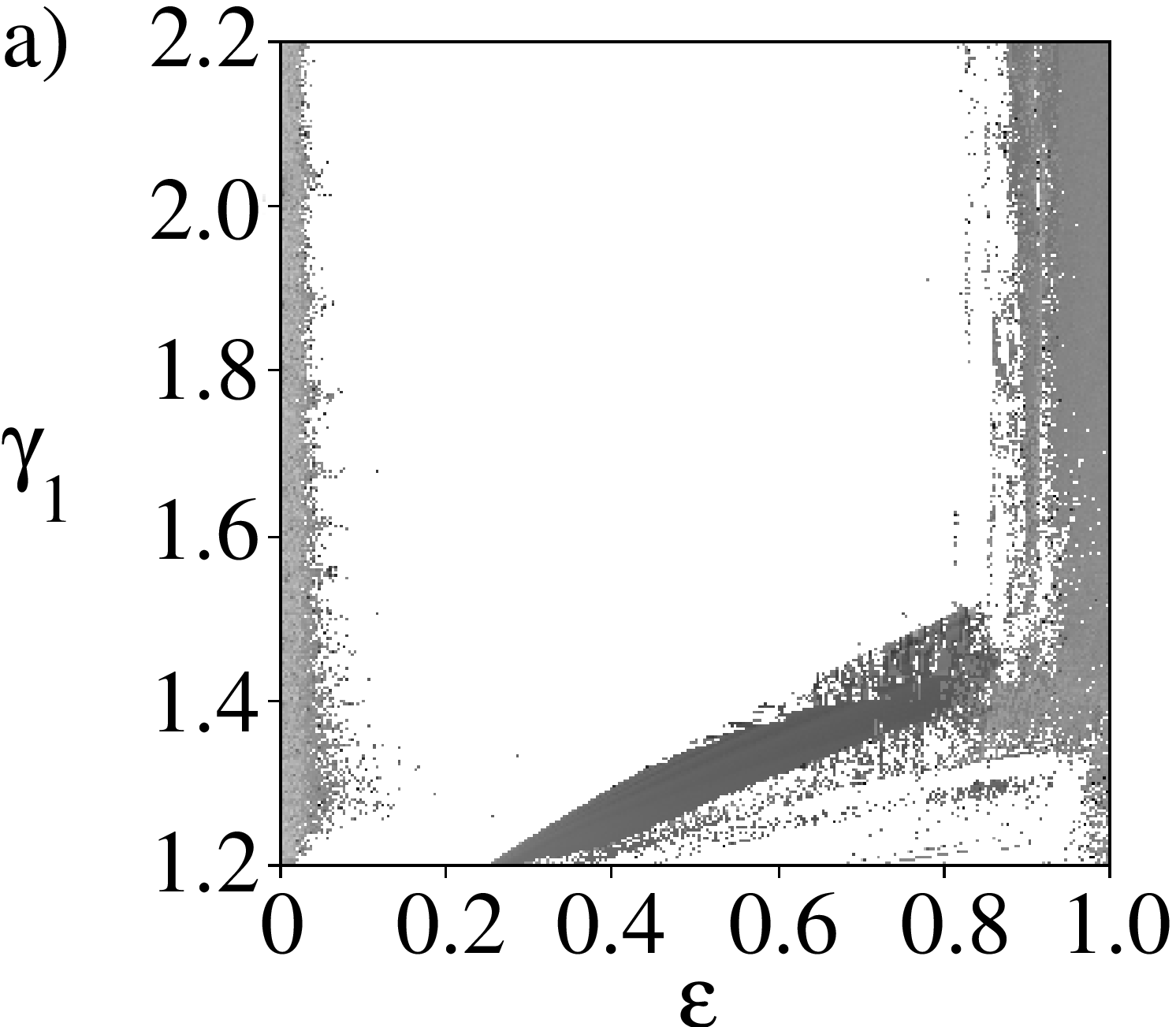}
\includegraphics[width=1.6in,angle=0]{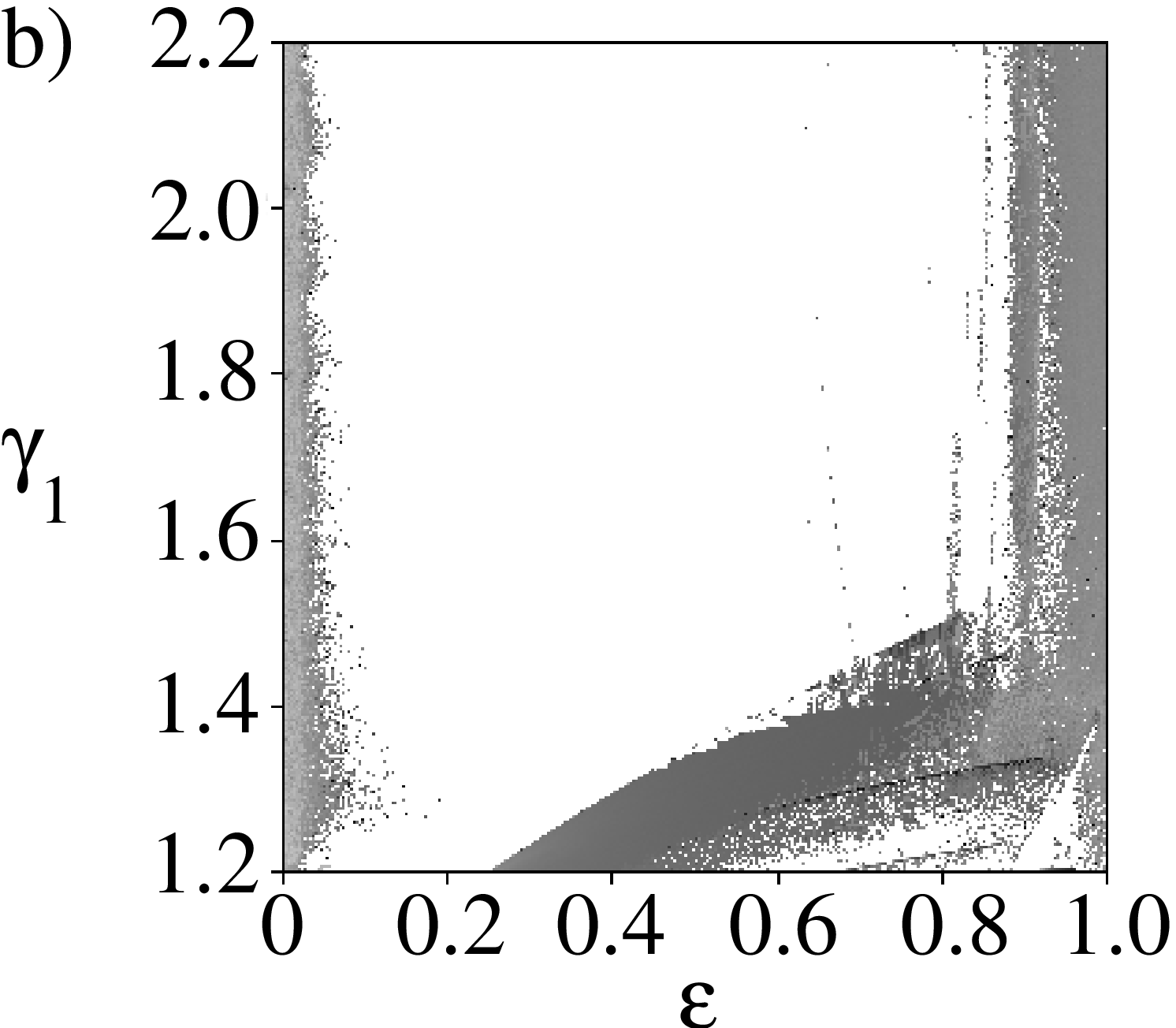}}}
\vspace{0.2cm}
\caption{The ring-type system with seven elements under the low-amplitude
spatially uniform periodic (a) and chaotic (b) fields $\xi_n=a y_n$, where
$a=0.0003$. 
The periodic field
is generated by $y_n = \sin \left( 2 \pi n / T \right)$ with $T=500$ and  
chaotic one by the logistic map $y_{n+1}=1-\lambda y_n^2$ 
with parameter $\lambda=1.99$.}
\label{fig7}
\end{figure}

The investigations of the synchronization also involved the possibility to 
influence on this process with an external signal. It was revealed that an 
external low-amplitude spatially uniform field (both periodical and chaotic) 
may increase the degree of synchronization in the ensemble. As the example 
of this fact  the parameter planes are presented  for ring-type system 
with seven neurons under an external force (Fig.~\ref{fig7}).

As one can see from comparison of the parameter plane in Fig.~\ref{fig7} and 
Fig.~\ref{fig6}a, the region of synchronous behaviour for the model neuron
ensemble under the external force becames larger in relation to the region
for unperturbated system.  

\section{Conclusion}

In this work the phenomenological map describing some aspects of neuron 
dynamics is presented. It is shown that the proposed map qualitatively 
simulates real neuron behavior and describes  some synchronization
phenomena that 
is observed in neuron ensembles, connected via electrical synapse. The 
assumption that just fast motion is responsible for the 
synchronization  phenomena is 
prior and should be confirmed. The results related to a possibility of 
low-amplitude external field to increase the degree of synchronization in 
small model neuron ensemble may be useful from the 
practical viewpoint but only in case 
that they will get an experimental verification.

The way on the creation of the piecewise map from phenomenological 
viewpoint (from view of  the system time series), 
applied in the work, could be used for other sysnems.  

This work was supported by RFBR under grants 02-02-16351, 00-15-96673, 
Program "Universities of Russia" and 
Ministry of Education of Russian Federation under 
grant E00-3.5-196.

\end{document}